\documentclass[prl,twocolumn,showpacs,superscriptaddress,aps]{revtex4-1}

\usepackage{amsmath}
\usepackage{amssymb}
\usepackage{graphicx}
\usepackage{bm}

\usepackage{color}
\usepackage{placeins}

\begin{document}

\title{Surface hopping methodology in laser-driven molecular dynamics}

\author{T.~Fiedlschuster}	
\affiliation{Institut f\"{u}r Theoretische Physik, Technische Universit\"{a}t Dresden, D-01062 Dresden, Germany}
\affiliation{Max-Planck-Institut f\"ur Mikrostrukturphysik, Weinberg 2, D-06120 Halle, Germany}

\author{J.~Handt}
\affiliation{Institut f\"{u}r Theoretische Physik, Technische Universit\"{a}t Dresden, D-01062 Dresden, Germany}

\author{E.~K.~U.~Gross}	
\affiliation{Max-Planck-Institut f\"ur Mikrostrukturphysik, Weinberg 2, D-06120 Halle, Germany}

\author{R.~Schmidt}
\email{Ruediger.Schmidt@tu-dresden.de}
\affiliation{Institut f\"{u}r Theoretische Physik, Technische Universit\"{a}t Dresden, D-01062 Dresden, Germany}

\date{\today}

\begin{abstract}
A theoretical justification of the empirical surface hopping method for the laser-driven molecular dynamics is given utilizing the formalism of the exact factorization of the molecular wavefunction [Abedi et al., PRL \textbf{105}, 123002 (2010)] in its quantum-classical limit. Employing an exactly solvable
$\textrm H_2^{\;+}$-like model system, it is shown that the deterministic classical nuclear motion on a single time-dependent surface in this approach describes the same physics as stochastic (hopping-induced) motion on several surfaces, provided Floquet surfaces are applied. Both quantum-classical methods do describe reasonably well the exact nuclear wavepacket dynamics for extremely different dissociation scenarios. Hopping schemes using Born-Oppenheimer surfaces or instantaneous Born-Oppenheimer surfaces fail completely.
\end{abstract}


\maketitle

For more than two decades, surface hopping (SH) \cite{tully_1990} has been among the most popular and successful methods to describe non-adiabatic phenomena in atomic many-body systems (for reviews see \cite{tully_2012,barbatti_2016,subotnik_2016,wang_2016}). From the theoretical point of view, however, any SH scheme is inherently a phenomenological approach. The ad hoc assumption of stochastic jumps between electronic potential energy surfaces (PES) has, so far, never been rigorously deduced from the time-dependent Schr\"odinger equation (TDSE) for electrons and nuclei, and even the choice of the applied PES is ambiguous.

Very recently, however, first attempts have been made to justify the SH methodology on Born-Oppenheimer surfaces (BOSs), solely for the laser-free non-adiabatic dynamics~\cite{abedi_2013,curchod_2013,subotnik_2013,kapral_2016}. A close similarity between the exact wavepacket propagation and SH on BOSs has been found in the framework of the exact factorization of the molecular wavefunction \cite{abedi_2013}. In this theory, the so-called exact time-dependent  potential energy surface (EPES),  together with an exact time-dependent vector potential, governs the true nuclear wavepacket dynamics. The EPES can exhibit nearly discontinuous step-like features, just in the vicinity of avoided crossings between BOSs, leading simultaneously to acceleration and deceleration of certain parts of the quantum wavepacket and resulting in its splitting. In close analogy, the SH mechanism can create branches of classical trajectories at avoided crossings. The findings \cite{abedi_2013} justify, albeit qualitatively but anyhow convincingly, the SH methodology on BOSs, in the field-free case.

For the laser-driven dynamics, any validation of SH is still lacking and the appropriate choice of the applicable PES is discussed controversially, at present \cite{mitric_2009,richter_2011,*richter_2012,suzuki_2015}. In fact, the hitherto purely intuitively chosen PES in SH models are fundamentally different from each other, and include BOSs \cite{jones_2008,mitric_2009,tavernelli_2010,mitric_2011,petersen_2012,bajo_2014}, instantaneous BOSs (IBOSs) \cite{dietrich_1996,thachuk_1996,thachuk_1998,kelkensberg_2011,bajo_2012} as well as Floquet surfaces (FSs) \cite{horenko_2001,bajo_2012,fiedlschuster_2016}. From the massive differences in definition and properties of these PESs, one can hardly expect that the appendant SH schemes can describe the same physics. Obviously, the situation requires clarification and the general questions persist: Is there any validation of SH methodology at all in this case, and if yes, what are the adequate PES?

In this paper, we will provide answers to both questions employing the quantum-classical limit of the exact factorization \cite{abedi_2010,abedi_2013,min_2014,min_2015} and using deliberately an exactly solvable model system. The exact factorization leads to a TDSE for the nuclear subsystem \emph{alone} which is exact in the sense that the absolute square of the corresponding, purely nuclear, wavefunction yields the exact nuclear N-body density of the full electron-nuclear system. Hence, if the true quantum-mechanical nuclear density is approximated by an ensemble of classical trajectories, the correct classical force on the nuclei is \emph{uniquely} given by the gradient of the EPES. In other words, the "classical" EPES contains all electron-nuclear correlations which generally can be retained in the quantum-classical limit of the TDSE. Consequently, ensembles of  classical trajectories on the EPES can serve as judge for all the phenomenological SH models. From extensive comparative numerical studies, it will become apparent that the role of the BOSs in the field-free case is taken over by FSs in the laser driven case, although the mechanism is more complex. Ensembles of classical trajectories, propagated stochastically on FSs (Floquet-SH, F-SH) and deterministically on the EPES (exact surface dynamics, ESD), do describe the same physics. Moreover, in the considered cases, the results are in excellent agreement with those of the TDSE.  Complementary SH calculations with BOSs (BO-SH) and IBOSs (IBO-SH) deliver unphysical results.

The soft-core center-of-mass Hamiltonian of the model system reads 

\begin{align}\label{eq:H_model}
H=&-\frac{\Delta_R}{2M}-\frac{\Delta_r}{2}+\frac{1}{R+0.03}-\mu F(t)cos(\omega t)\nonumber\\
&-\frac{1}{\sqrt{(r-R/2)^2+1}}-\frac{1}{\sqrt{(r+R/2)^2+1}},
\end{align} 

with $M$ the reduced nuclear mass, $R$ the internuclear distance, and $r$ the electronic position operator. The interaction of the molecule with the linear polarized laser (carrier frequency $\omega=0.2$ a.u. [$\lambda\approx 225$ nm] throughout the whole paper) is included in the length gauge (dipole operator $\mu=-r$). We apply laser fields where the envelope $F(t)$ does not change considerably during one optical cycle $T=2\pi/\omega\approx0.8$ fs. Atomic units are used unless stated otherwise.

\begin{figure}[h]
\begin{center}
\vspace{0.5cm}
\includegraphics[scale=0.4]{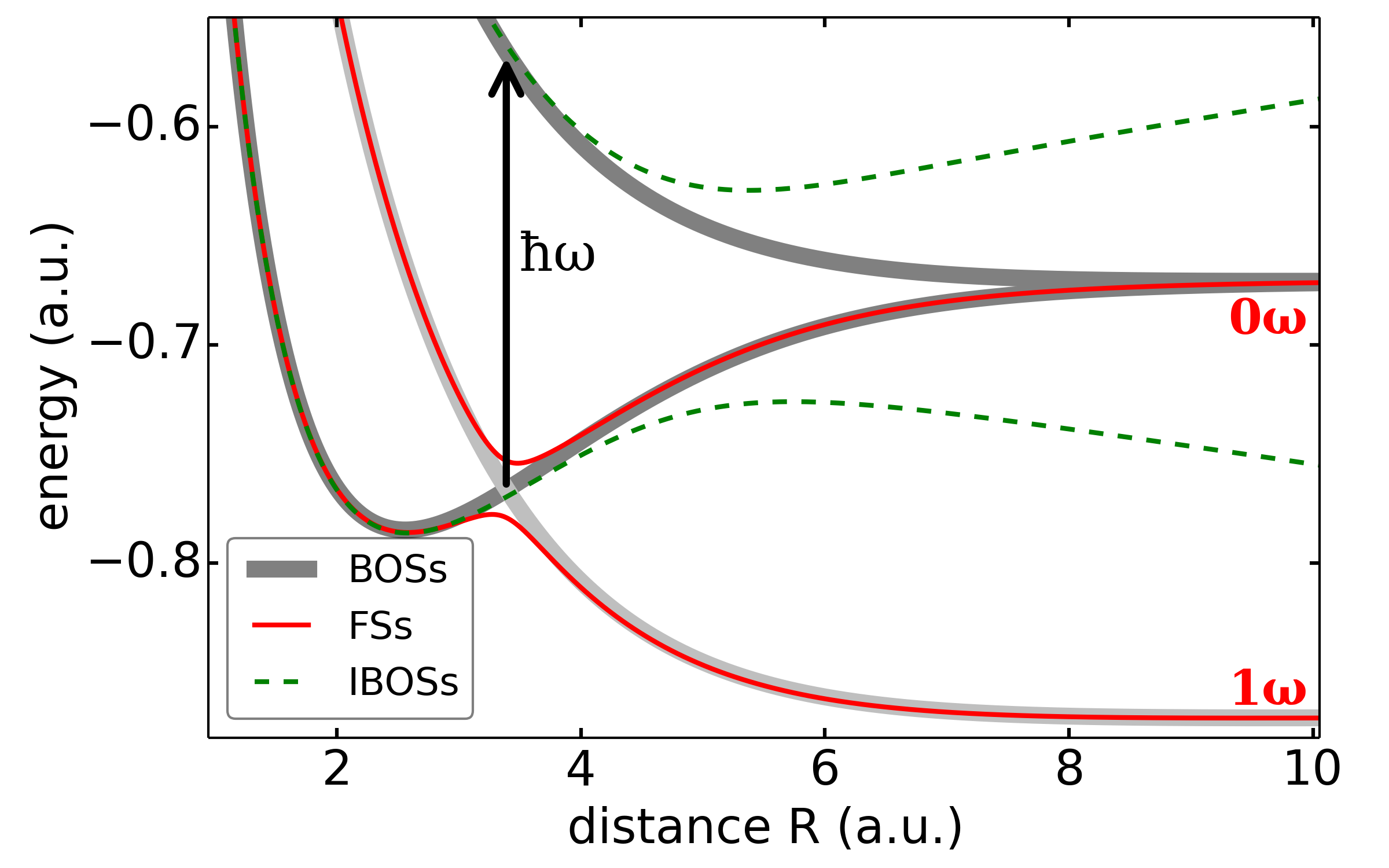}
\caption{The two lowest BOSs and IBOSs (for $\left|\cos \omega t\right|=1$), the excited BOS shifted down by $\hbar \omega$ (light gray), as well as the two FSs ($0\omega$ and $1\omega$ FS) forming the one-photon crossing for a $\lambda=225$~nm laser with $I=10^{13}$ W/cm${}^2$.  The vertical arrow of length $\hbar\omega$ marks the one-photon resonance between the BOSs.} 
\label{fig:fig1}
\end{center}
\end{figure}     

The various PES (BOSs, IBOSs, and FSs), for the model system \eqref{eq:H_model} are presented in Fig.~\ref{fig:fig1} for a laser intensity I=$10^{13}$ W/cm${}^2$. For clarity, only the two lowest BOSs and IBOSs as well as two (relevant) FSs are plotted. The different time dependence of the PES is visualized in the supplemental material (SM) \cite{sm}.
The displayed PES look very different, although they do have some common aspects: The IBOSs are identical with the BOSs whenever $\cos \omega t=0$. The FSs are piece-wise equal to BOSs, appropriately shifted  by the photon energy $\hbar \omega$ (dressed BOSs). The decisive difference between the PES concerns their behavior just at the one photon resonance located at $R\approx3.5$  a.u.
Whereas BOSs and IBOSs do not show any peculiarities, the FSs exhibit an avoided crossing. These typical crossings are the crucial difference to all other PES. The resulting gap size (even tunable by the electric field strength) allows for both, deterministic evolution on one FS surface (without hops) or stochastic dynamics on both FS (with hops). To what extent this peculiarity favors the use of FSs in corresponding SH schemes will be analyzed in the following comprehensive dynamical calculations.

In these calculations, we will consider the most detailed observable quantity of all, namely the resulting nuclear probability density distributions (NPDDs) in position and momentum space. This allows for a direct comparison with the exact wavepacket dynamics of the TDSE, and excludes artificial agreement between the different methods in (possibly) insensitive integral quantities. For the classical initial conditions, we use the  Wigner distributions of the exact initial quantum wavepackets, details are given in the SM \cite{sm}. 
The various methods to be applied are outlined in detail in previous publications (for BO-SH see \cite{fischer_2014,fischer_2014a,fischer_2014b} \footnote{Decisive arguments against the use of BO-SH for laser-driven molecular dynamics have been discussed already in \cite{fischer_2014,fischer_2014b}, albeit qualitatively. Here we use intentionally the general BO-SH formalism of \cite{fischer_2014} for the laser-driven case to prove the arguments quantitatively.}, for IBO-SH see \cite{thachuk_1996,thachuk_1998}, for F-SH see \cite{fiedlschuster_2016}, all three methods are based on Tully's fewest switching algorithm \cite{tully_1990}). To make this paper self-contained, the SM \cite{sm} includes a brief summary of the hopping methods and of the calculation of the EPES from the solution of the TDSE (see also \cite{abedi_2010,abedi_2012}).  In the following, we will consider various dissociation scenarios with initial conditions which ensure extremely different mechanisms.

\paragraph{Scenario 1 (photon absorption)}

\begin{figure}[h]
\begin{center}
\includegraphics[scale=0.4]{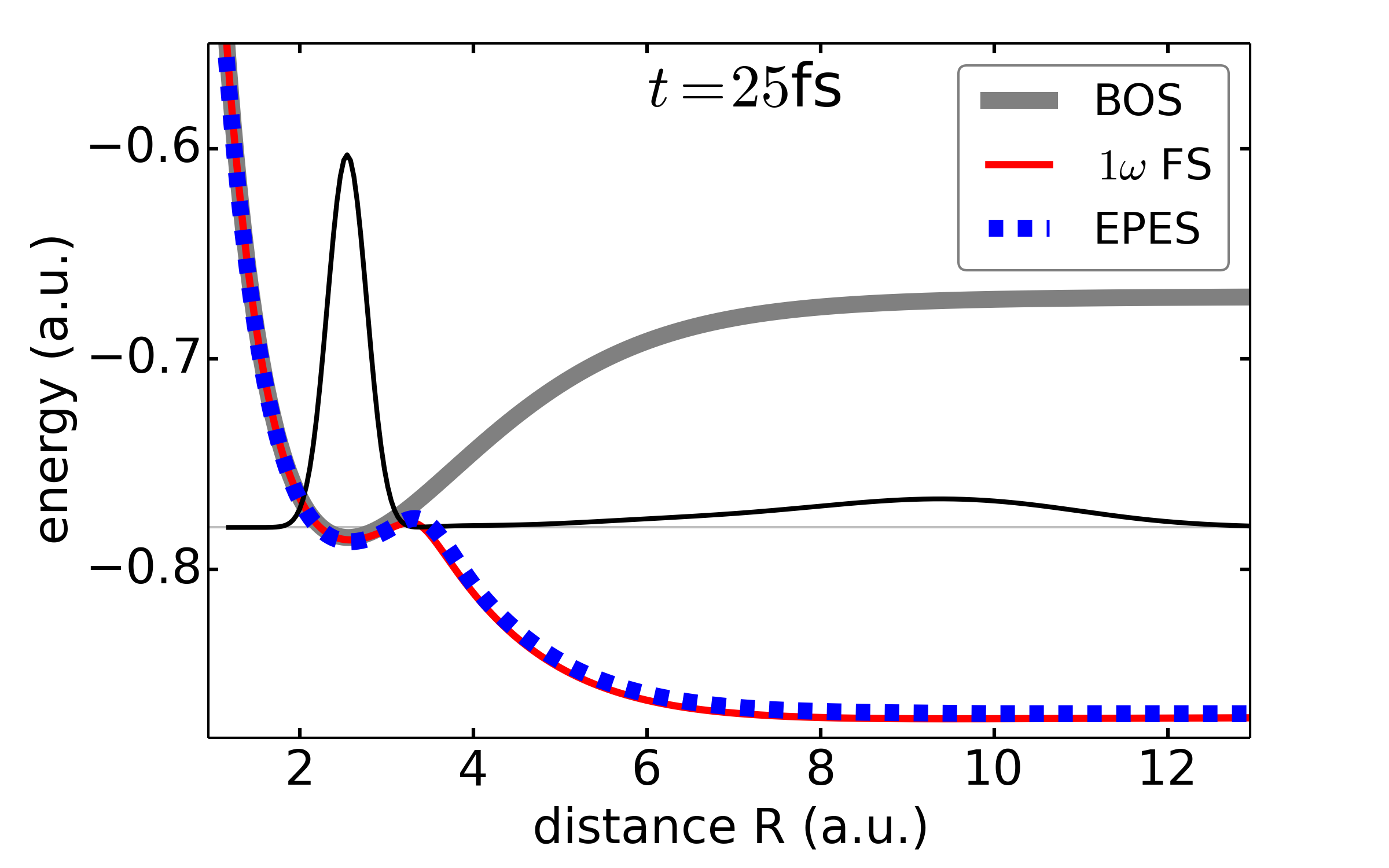}
\caption{Wavepacket of the TDSE for scenario~1 (thin black line), together with the groundstate BOS, the $1\omega$ FS and the EPES  at $t=25$~fs.} 
\label{fig:fig2}
\end{center}
\end{figure}

\begin{figure}[h]
\begin{center}
\includegraphics[scale=0.3]{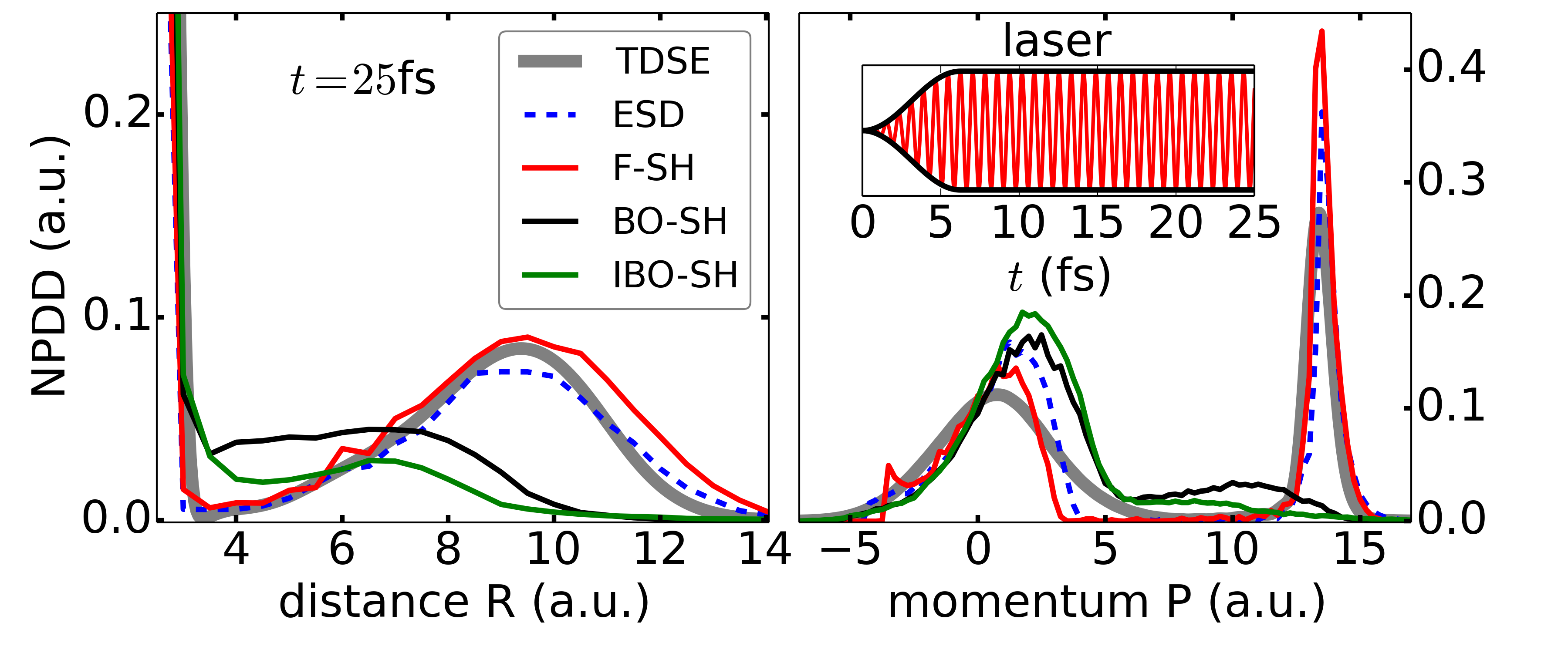}
\caption{The NPDDs for scenario 1 at $t=25$~fs, obtained by solving the TDSE and with the different quantum-classical methods, in position (left panel) and momentum space (right panel). In the inset, the electric field and the envelope of the laser are shown.} 
\label{fig:fig3}
\end{center}
\end{figure}

The $\textrm H_2^{\;+}$-like molecule with $M=918$ a.u. is initially in its groundstate with an additional momentum of $-2.5$ a.u. applied to the nuclei (see SM \cite{sm}). It is exposed to a cw-laser with $I=10^{13}$ W/cm${}^2$\; (as used for Fig.~\ref{fig:fig1}), switched on with a $\sin^2$-shaped ramp (see inset in Fig.~\ref{fig:fig3}). In Fig.~\ref{fig:fig2}, the resulting wavepacket of the TDSE at $t=25$ fs is shown. Most of the initial wavepacket remains bound, localized at the equilibrium distance at $R\approx 2.5$ a.u. The dissociating part exhibits a maximum at $R\approx 9$ a.u. In Fig.~\ref{fig:fig3}, it is presented together with the momentum distribution, where the dissociating part is sharply localized at $P\approx 13.5$ a.u. The corresponding  kinetic energy release of the fragments of $P^2/2M\approx 0.1$ a.u. equals the difference between the photon energy $\hbar \omega=0.2$ a.u. and the binding energy of the molecule in its vibrational groundstate of $E_0\approx0.1$ a.u. This perfect energy balance strongly suggests (although does not conclusively prove) that one photon absorption is the dominant dissociation mechanism.

The interesting question now is, how the EPES of this scenario, calculated from the exact solution of the TDSE, compares to the different PES discussed above and presented in Fig.~\ref{fig:fig1}. In Fig.~\ref{fig:fig2}, the groundstate BOS, the $1\omega$ FS and the EPES are shown at $t=25$ fs. The EPES is averaged over one optical cycle of the laser, the whole time dependence of  the EPES and the $1 \omega$ FS is visualized in the SM \cite{sm}. Evidently, and indeed surprisingly, the averaged EPES and the $1 \omega$ FS coincide perfectly at all distances and all times! Hence, the deterministic quantum-classical dynamics on both surfaces is definitely the same. To a large extent, this should hold also for an ensemble of trajectories in ESD and full F-SH calculations, as long as the number of hops per trajectory $N$ between FS remains very small ($N \ll 1$). This is indeed the case (see Tab.~\ref{tab:tab1}) and can also be expected from the discussion of Fig.~\ref{fig:fig1}. Hence, and now not surprisingly, the NPDDs obtained with ESD and full F-SH calculations are nearly equal (see Fig.~\ref{fig:fig3}). In addition, they do compare nicely with that of the TDSE in position as well as in momentum space. 

In contrast, the NPDDs calculated with BO-SH and IBO-SH are in qualitative
disagreement with that obtained by the other methods (see Fig.~\ref{fig:fig3}). Any dissociation of the molecule on these surfaces requires stringently a certain number of hops (see Fig.~\ref{fig:fig1}), which do occur (see Tab.~\ref{tab:tab1}) but, at the same time, lead to fundamentally different nuclear dynamics.

Summarizing this part, it was found, somewhat surprisingly, that the EPES can coincide with a single FS. In this case, the non-adiabatic dynamics proceeds deterministically, i.e., without any hops in the F-SH procedure. In the following we will consider a scenario where hops between FSs stringently do occur.
\paragraph{Scenario 2 (photon emission)}

\begin{figure}[h]
\begin{center}
\includegraphics[scale=0.3]{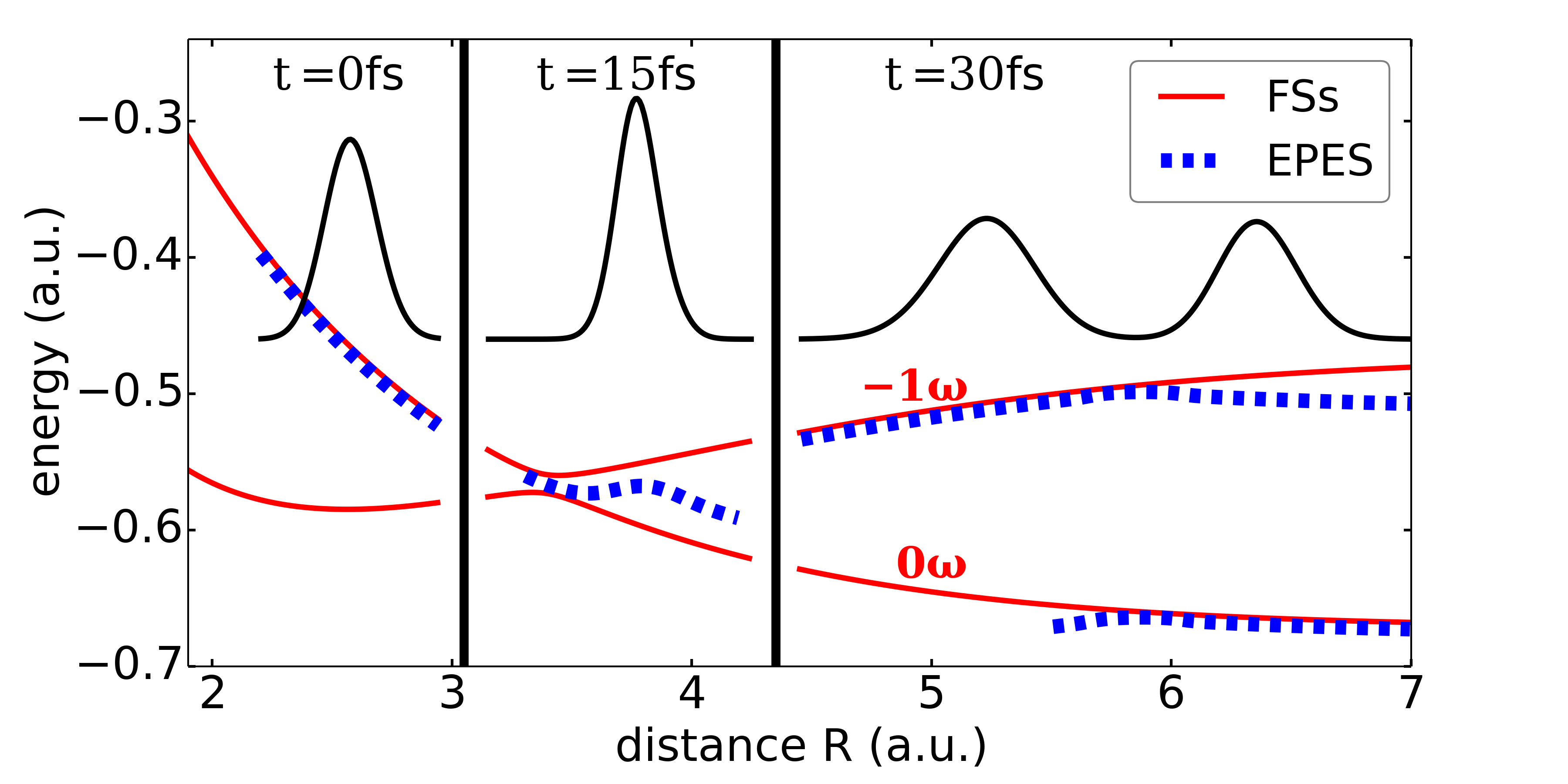}
\caption{Wavepacket of the TDSE for scenario 2 (snapshots at $t=0$, $15$ and $30$~fs; thin black lines with different scales), together with the relevant FSs and the EPES. For $R > 5.5$~a.u., the EPES is also plotted shifted down to the $0\omega$~FS.} 
\label{fig:fig4}
\end{center}
\end{figure}

\begin{figure}[h]
\begin{center}
\includegraphics[scale=0.3]{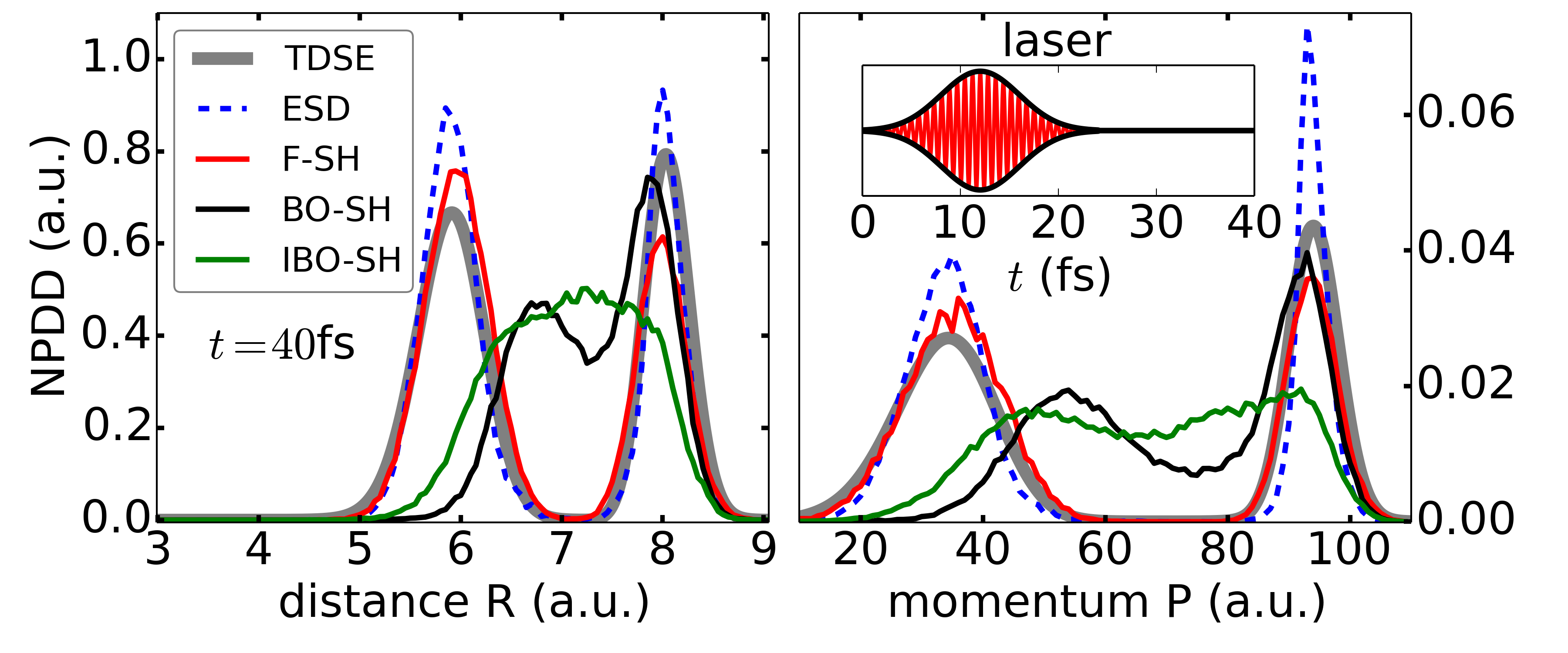}
\caption{The NPDDs for scenario 2 at $t=40$~fs, obtained by solving the TDSE and with the different quantum-classical methods, in position (left panel) and momentum space (right panel). In the inset, the electric field and the envelope of the laser pulse are shown. } 
\label{fig:fig5}
\end{center}
\end{figure}

A $\textrm Na_2^{\;+}$-like molecule ($M=23\times 918$ a.u.) is initially lifted into its first excited electronic state with a nuclear wavefunction of its (Franck-Condon projected) vibrational groundstate (see wavepacket at $t=0$ in Fig.~\ref{fig:fig4}). It is exposed to a short Gaussian-shaped laser pulse of $7$ fs FWHM, wavelength of $\lambda=225$ nm, and peak intensity of $I=3\times10^{12}$ W/cm${}^2$ (see inset in Fig.~\ref{fig:fig5}). For this scenario, the wavepacket dynamics resulting from the TDSE is depicted at different times $t=0,15,30$~fs in Fig.~\ref{fig:fig4}. The final NPDDs of the TDSE in position and momentum space are given in Fig.~\ref{fig:fig5}, at $t=40$~fs. As clearly seen, the initial wavepacket is split into a fast-moving part (with mean momentum of $P \approx 95$~a.u.) and a slow-moving part (with $P \approx 35$~a.u.). The mean kinetic energy  of the fast one $P^2/2M\approx0.21$~a.u. corresponds to the energy difference on the excited BOS between the initial mean internuclear distance of $R \approx 2.5$~a.u. ($E\approx-0.46$~a.u.) and the final one of $R \approx 8$~a.u ($E\approx-0.67$~a.u.), reflecting free motion (sliding down) on this surface as dissociation mechanism. On the other hand, the mean kinetic energy of the retarded part $P^2/2M\approx0.03$~a.u. is smaller by almost one photon energy $\hbar \omega=0.2$~a.u.,  suggesting (but not definitely proving) stimulated photon emission as the dissociation mechanism for this fraction of the wavepacket.

In Fig.~\ref{fig:fig4}, snapshots of the corresponding EPES as well as of the $0 \omega$ and $-1 \omega$~FSs are presented (for the time-evolution see SM \cite{sm}). Field-free FSs are generally identical with  BOSs, dressed by a certain number of photons. The EPES, respectively its gradient, also coincides with BOSs, in the field-free case~\cite{abedi_2013}. Thus, at $t=0$ the first excited BOS is equal to the EPES as well as the $-1 \omega$~FS (in our notation). The $0 \omega$~FS equals the groundstate BOS, dressed (shifted up) by one photon. After the pulse at $t=30$~fs, the FSs changes their assignment with respect to the (dressed) BOSs. The EPES, however, and indeed somewhat surprisingly, coincides with the (one photon shifted) groundstate BOS (and the $-1 \omega$~FS) in the range of the retarded part of the wavepacket, and with the excited BOS (and the $0 \omega$~FS) in the region of the fast moving part. We note in passing that this already proves the interpretation of the dissociation mechanisms for both parts of the wavepacket given above.

During the laser pulse ($t=15$ fs in Fig.~\ref{fig:fig4}), the EPES does not coincide with one of the other surfaces. Its alternating gradients lead, at the same time, to acceleration and deceleration of certain parts of the wavepacket of the TDSE, resulting in the splitting of the wavepacket. This pure quantum mechanical effect survives the crude quantum-classical approximation in terms of decelerated and accelerated classical trajectories in appendant ESD calculations, which is convincingly demonstrated in Fig.~\ref{fig:fig5}. The NPDDs, resulting from the deterministic ESD calculations, are in excellent agreement with that of the TDSE in position as well as in momentum space. 

As an important result of this work, it will be shown in the following that the same mechanism can be clearly understood and adequately described also with the stochastic SH methodology, provided FSs are applied. In Fig.~\ref{fig:fig4}, both relevant FSs are shown at $t=15$ fs. They exhibit a typical  avoided crossing located at the one photon resonance ($R \approx3$~a.u.). This crossing induces a strong non-adiabatic coupling between both surfaces. Thus, in classical F-SH calculations, trajectories staying on the upper FS are decelerated owing to the loss of one photon. Trajectories performing one hop between the upper and lower FS are further accelerated and slide down the initial excited BOS. The almost equal partition of both types of trajectories (see number of hops per trajectory in Tab.~\ref{tab:tab1}) will lead to an almost symmetric splitting of the NPDD in position space with two pronounced maxima in momentum space.  The results of the dynamical F-SH calculations are in intriguing agreement with ESD calculations as well as the  TDSE solution (see Fig.~\ref{fig:fig5}).

Avoided crossings, located at photon resonances, basically do not exist between BOSs and IBOSs, owing to their photon-less definition.  On the other side, it is just the additional non-adiabatic coupling $\sim \dot R \langle \Phi_{-1\omega}|\partial_R \Phi_{0\omega}\rangle$ between the Floquet states $|\Phi\rangle$, which leads to the different dissociation mechanisms. Hence, it is not very surprising that the SH calculations with BOSs or IBOSs yield NPDDs which disagree, even qualitatively, with that of F-SH (see Fig.~\ref{fig:fig5}). In these approaches, hops are (mainly) created by the laser-induced coupling ($\sim R F(t)\cos \omega t$ for BOSs~\cite{fiedlschuster_2016}, see~\cite{thachuk_1996} for IBOSs) and do occur at all internuclear distances. Accordingly, the number of hops is distinctly larger than in the F-SH approach (middle column in Tab.~\ref{tab:tab1}).

\begin{figure}[h]
\begin{center}
\includegraphics[scale=0.3]{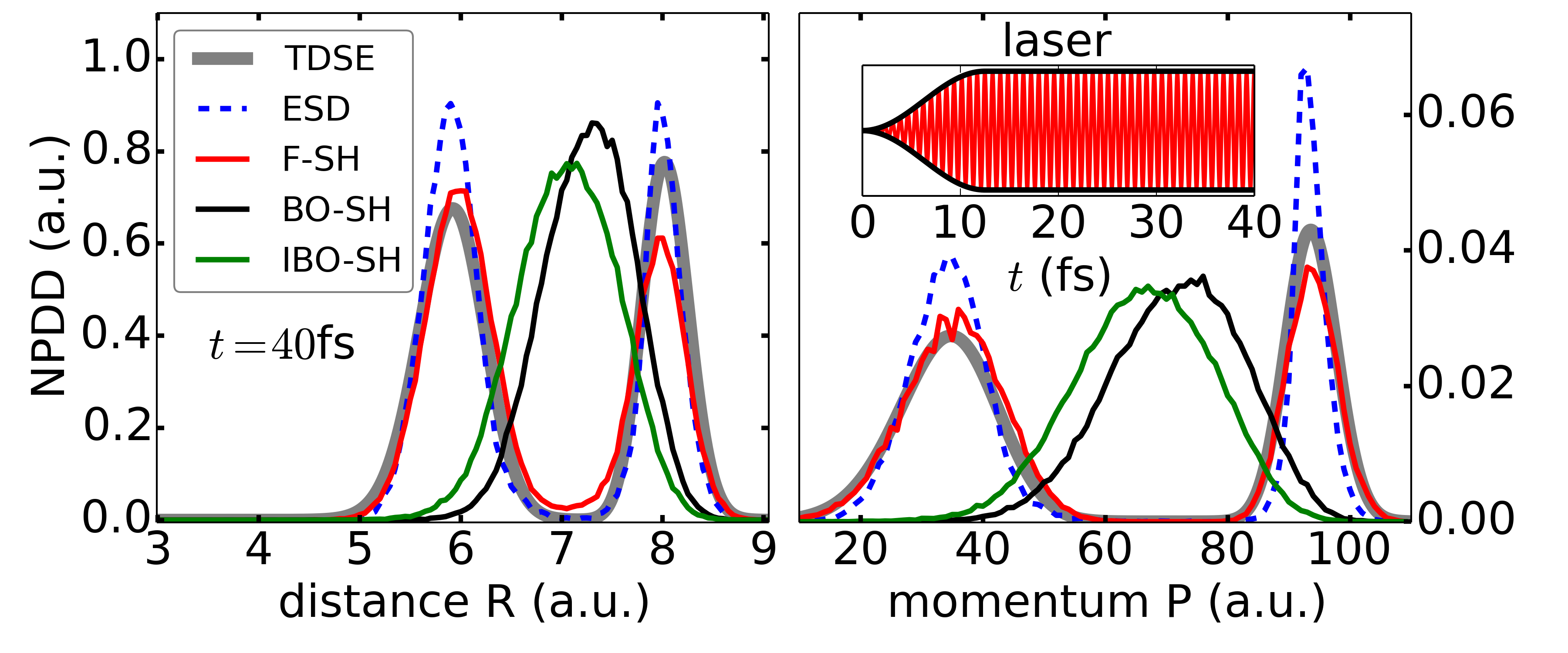}
\caption{The NPDDs for scenario 2 at $t=40$~fs, using a cw-like laser (see inset) instead of a finite pulse.} 
\label{fig:fig6}
\end{center}
\end{figure}     

To further examine the different SH methods, we repeat the dynamical calculations of scenario~2, but replace the short laser pulse by a cw-like laser (see inset in Fig.~\ref{fig:fig6}). From the discussion above, the results of the F-SH calculations are expected to remain largely unaffected by this change, because the whole dissociation process is determined during a short time interval of about $15$~fs where both laser fields are practically equal (cf. insets in Figs.~\ref{fig:fig5} and~\ref{fig:fig6}). Indeed, the calculated NPDDs are nearly identical with that obtained for the short laser pulse (cf.  NPDDs in Figs.~\ref{fig:fig5} and~\ref{fig:fig6}). In addition, they do agree nicely with that of the TDSE and ESD.  In striking contrast, the NPDDs calculated with BO-SH or IBO-SH for the cw-like laser are drastically different from that obtained for the short pulse (cf. Figs.~\ref{fig:fig5} and~\ref{fig:fig6}). This unphysical behavior results from the artificial large number of hops (see last row in Tab.~\ref{tab:tab1}) which, in addition, will further increase in time (up to $N\to \infty$!) in both approaches.

\begin{table}[h]
\begin{center}
\begin{ruledtabular}
\begin{tabular}{llclclc}
method& & scenario 1 & & scenario 2 (pulse) & & scenario 2 (cw) \\
\hline
BO-SH & &2.6 & &1.0 & &15.7 \\
IBO-SH & &2.9 & &2.5 & &55.9 \\
F-SH & &0.01 & &0.6 & &0.7
\end{tabular}
\end{ruledtabular}
\caption{The average number of hops/trajectory in the different surface hopping methods for the investigated scenarios.} 
\label{tab:tab1}
\end{center}
\end{table}

In summary,
we have performed a comprehensive study of the laser-driven dynamics, using different quantum-classical approximations of the TDSE for electrons and nuclei.
We have shown that the inherently deterministic propagation of the nuclei on the EPES can be equivalently
described by stochastic motion on several surfaces mediated by hops between them,
provided FSs are used.
Both methods (ESD and F-SH) deliver the same results which, in addition, are in excellent agreement with the exact wavepacket dynamics of the full electron-nuclear TDSE.
The studies justify the SH methodology for the laser-driven case.
At the same time, the investigations also conclude the present, controversially led debate about the applicability of BOSs or IBOSs in SH schemes~\cite{mitric_2009,richter_2011,*richter_2012,suzuki_2015},
because both are not appropriate.
These conclusions are valid for (and at the same time limited to) laser fields where the Floquet treatment of the time-dependent Hamiltonian approximately applies. 

Whereas the solution of the full electron-nuclear TDSE is restricted to small model systems, the SH approach can be applied to realistic systems and should, as we have shown, reproduce the correct laser-driven dynamics as long as the stochastic hopping is done between Floquet surfaces.

\begin{acknowledgments}
We gratefully acknowledge financial support from the Deutsche Forschungsgemeinschaft through the Normalverfahren (Nr. SCHM 957/10-1).
\end{acknowledgments}

\FloatBarrier

%\bibliography{literatur_tobias}

\end{document}